\documentclass[preprint,12pt,authoryear]{elsarticle}

\usepackage[pdftex]{color,hyperref}
\usepackage{amsmath,amsfonts,amssymb}
\usepackage{fullpage}
\usepackage[caption=false]{subfig}
\usepackage{float}
\usepackage{mathscinet}

\newcommand{\eref}[1]{Eq.~(\ref{#1})}
\newcommand{\esref}[2]{Eqs.~(\ref{#1})-(\ref{#2})}
\newcommand{\sref}[1]{Section~\ref{#1}}
\newcommand{\ssref}[2]{Sections~\ref{#1}-\ref{#2}}
\newcommand{\fref}[1]{Fig.~\ref{#1}}

\journal{Handbook of Computational Social Science}

\begin{document}

\begin{frontmatter}

\title{Opinion dynamics in social networks: From models to data\tnoteref{t1}}
\tnotetext[t1]{This is a draft chapter. The final version will be available in {\it Handbook of Computational Social Science} edited by Taha Yasseri, forthcoming 2023, Edward Elgar Publishing Ltd. The material cannot be used for any other purpose without further permission of the publisher, and is for private use only.
Please cite as: Peralta, A. F., Kert\'esz, J., and I\~{n}iguez, G. (2023). Opinion dynamics in social networks: From models to data. In: T. Yasseri (Ed.), {\it Handbook of Computational Social Science}. Edward Elgar Publishing Ltd.}

\author[1]{Antonio F. Peralta}
\author[1,2]{J\'anos Kert\'esz}
\author[1,3,4]{Gerardo I\~{n}iguez\corref{cor1}}
\address[1]{Department of Network and Data Science, Central European University, A-1100 Vienna, Austria}
\address[2]{Complexity Science Hub, A-1080 Vienna, Austria}
\address[3]{Department of Computer Science, Aalto University School of Science, FI-00076 Aalto, Finland}
\address[4]{Centro de Ciencias de la Complejidad, Universidad Nacional Auton\'{o}ma de M\'{e}xico, 04510 Ciudad de M\'{e}xico, Mexico}
\cortext[cor1]{Corresponding author}
\ead{peraltaaf@ceu.edu, kerteszj@ceu.edu, iniguezg@ceu.edu}

\begin{abstract}
\footnotesize
Opinions are an integral part of how we perceive the world and each other. They shape collective action, playing a role in democratic processes, the evolution of norms, and cultural change. For decades, researchers in the social and natural sciences have tried to describe how shifting individual perspectives and social exchange lead to archetypal states of public opinion like consensus and polarization. Here we review some of the many contributions to the field, focusing both on idealized models of opinion dynamics, and attempts at validating them with observational data and controlled sociological experiments. By further closing the gap between models and data, these efforts may help us understand how to face current challenges that require the agreement of large groups of people in complex scenarios, such as economic inequality, climate change, and the ongoing fracture of the sociopolitical landscape.
\end{abstract}

\begin{keyword}
{\footnotesize
Opinion dynamics; Social networks; Mathematical modeling; Computerized experiments
}
\end{keyword}

\end{frontmatter}

\section{Introduction}
\label{sec_intro}

We are surrounded by opinions, by pictures in our heads \citep{lippmann1992public}. Whether it's about politics, art, or each other, we constantly engage with views on a broad diversity of issues, and these opinions strongly influence our actions and social behavior. By processing new information and interacting with others, opinions may change in time, contributing to the evolution of the public debate and society itself. A search for the individual and collective mechanisms driving the dynamics of opinion is crucial for understanding the rise of social movements and large-scale societal change. The study of opinion dynamics also plays a central role in several fields of application including politics, governance and marketing. During the last decades, considerable attention has been devoted to measure and model the way we express, share, and change opinions \citep{Castellano:2009,xia2011opinion,holme2015mechanistic,Sirbu:2017,Jedrzejewski:2019}: How do people reach consensus, if any? What is the role of the underlying social network of interactions? How do opinions polarize and fragment within social groups? What is the role of media, government and the industry? How do the algorithms behind online social platforms influence our opinions? and so on. 

Computational social science \citep{lazer2009life,conte2012manifesto,wagner2021measuring} is a new approach to understand the structure and dynamics of human societies, including opinion dynamics, where computational algorithms mix the theories of sociology, economics, and political science with the quantitative rigor of applied mathematics and physics. Research in this emerging discipline lies in a spectrum from data- to model-driven: One end focuses on the acquisition and exploration of large-scale social data, particularly in online communication platforms \citep{gonzalez2013social}; the other deals with the simulation and analysis of multi-agent models with simplified rules of interaction \citep{holme2015mechanistic}. Ideally, these approaches should complement each other and provide information about opinion dynamics at different scales. 

Collecting empirical data on opinions is hard: we need to turn something as ambiguous and multifaceted as an opinion into a quantitative measure open to modeling, and then convince people to express it out loud. Even when opinion is reduced to a few choices, like in sociological surveys and questionnaires, data gathering efforts are limited by sample size and answer subjectivity \citep{kitchenham2008personal,berinsky2017measuring}. A notable exception are elections and the polls preceding them, which provide population-level data on political candidates of choice, as well as the motivations and attitudes behind voters' actions \citep{merrill1999unified,easley2010networks}. Another recent source of opinion data is the output of sentiment analysis and other natural language processing tools in online social media and recommendation engines \citep{liu2012sentiment}. Still, the sparsity of data has made modelling and simulation the predominant methods of the field, where mechanisms coming from sociological theory (or otherwise considered reasonable) are turned into mathematical rules and analyzed, often referring to common sense instead of empirical validation \citep{Castellano:2009,sobkowicz2009modelling}.

Opinion dynamics models inspired by statistical physics and mathematics go back at least 50 years to the sociodynamics model of Weidlich \citeyearpar{weidlich1971statistical} and the famous voter model introduced by Holley and Liggett \citeyearpar{Holley:1975} \citep{Redner:2019}. These first attempts at emulating the emergence of collective opinion from individual action were followed by a multitude of models with endless variations. In most of them, individuals are represented by agents who share opinions with their neighbors in an underlying social network, which may or may not evolve in time. Agents are characterized by an opinion state variable, discrete or continuous and uni- or higher-dimensional, depending on the scenario of interest. Due to the interactions between agents (pairwise or group-based), opinion state variables change in time, leading to collective states of consensus, polarization, or fragmentation of opinion, closely related to the topology of the network underneath \citep{Castellano:2009,xia2011opinion}.

In spite of the limited accessibility of empirical data to compare with, the study of opinion dynamics models provides insight into the social mechanisms behind information exchange, public opinion, and group formation. The versatility of the models allows us to study a variety of effects, ranging from media influence (e.g., an external field affecting all agents \citep{Sirbu:2017}) and algorithmic bias \citep{bozdag2013bias,Peralta:2021a}, to the adaptive interplay between opinion dynamics and social network structure \citep{gross2008adaptive,iniguez2009opinion}. A number of methods of analysis have been developed, from exact and approximate analytical treatments to efficient simulation schemes \citep{porter2016dynamical,laurence2018exact}. There are mathematical and conceptual relations between opinion dynamics and processes of epidemic spreading \citep{Satorras:2015} and social contagion \citep{lehmann2018complex,zhang2019empirically}, leading to a fruitful exchange of methods between the fields. Recently, computerized social network experiments with human subjects (sometimes together with bots \citep{stewart2019information}) have started to explore the extent to which simple rules of interaction emulate the actual dynamics of opinion expression, at least in controlled settings \citep{chacoma2015opinion,vande2016modelling}.

In this short and by no means exhaustive review of opinion dynamics in social networks, we go through some of the most well-known models in the literature, classified into models with discrete or continuous opinions (\sref{sec_models}). We also review attempts at validating idealized mechanisms of opinion evolution with either observational data (particularly in elections and polls) or controlled experiments (\sref{sec_data}). We conclude by summarizing directions in which opinion dynamics models are becoming more realistic (\sref{sec_structural}), alongside an outlook on the relevance and future of the field (\sref{sec_conclusions}).

\section{Models of opinion dynamics}
\label{sec_models}

Modeling opinion exchange typically means proposing rules for the ways in which individuals interact and influence each other, alongside a set of features or states that characterize people's opinions in a specific social context. Opinion dynamics models can be classified as discrete or continuous depending on the nature of this opinion state variable. In discrete models, the opinion of an individual is one of a finite number of possibilities, such as a political candidate, a choice of product, or a pre-defined answer in a survey. Continuous models, on the other hand, use real-valued state variables (usually in an interval) to capture the level of (dis-)agreement with an opinion, attitude, or topic of interest.

\subsection{Models with discrete opinions}
\label{sec_discrete_models}
    
\begin{figure}[t]
\centering
\includegraphics[width=0.85\textwidth]{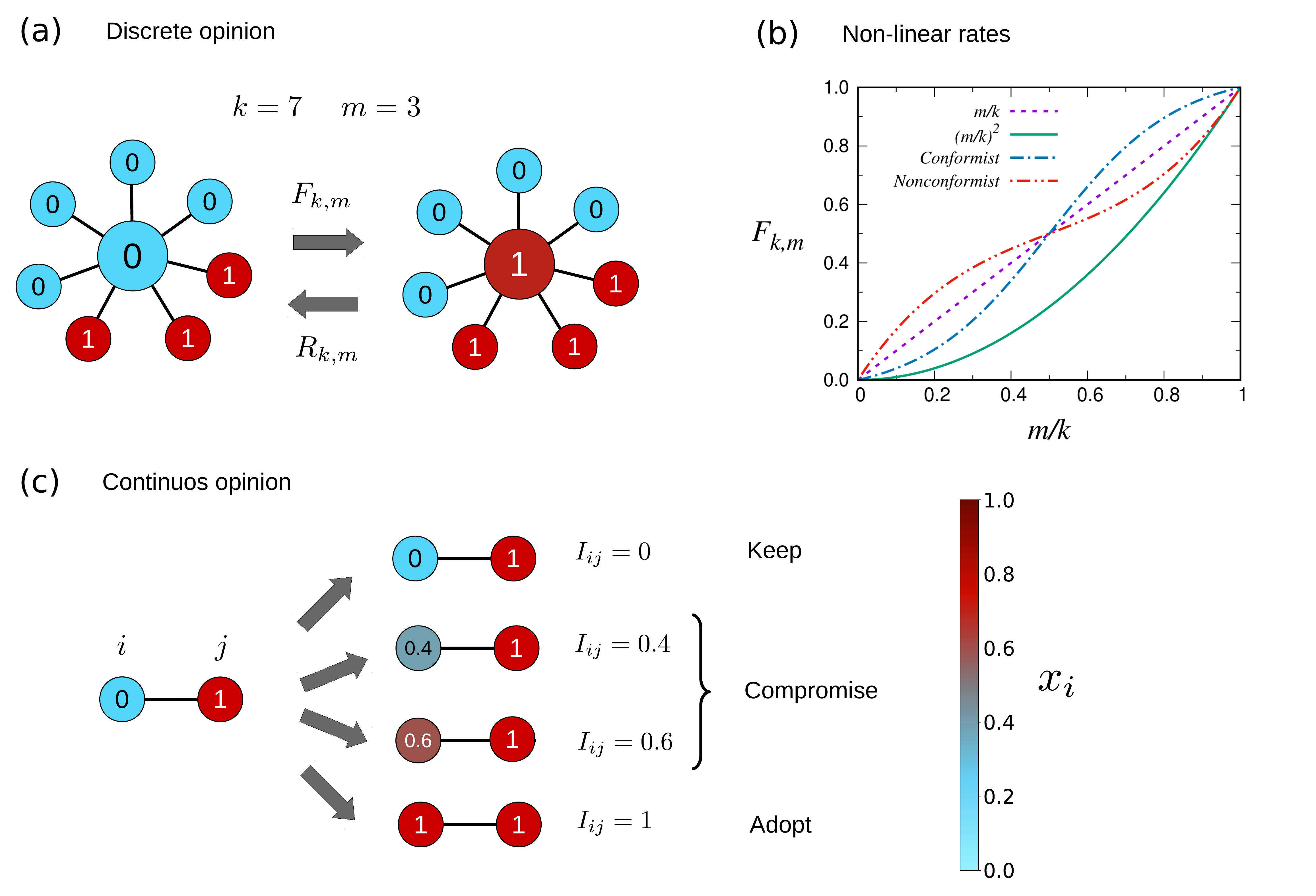}
\caption{\footnotesize {\bf Opinion dynamics models with discrete/continuous opinions.} {\bf (a)} In the case of two opinions (represented as variables $x = 0, 1$, or contrasting colors), transition rates between states ($F_{k,m}$ for infection, $x = 0 \rightarrow 1$; and $R_{k,m}$ for recovery, $x = 1 \rightarrow 0$) can be written as a function of network degree $k$ and the number of neighbors $m$ in state 1 \citep{Gleeson:2011}. {\bf (b)} Examples of infection rate $F_{k,m}$ (in terms of the fraction of neighbors $m/k$ holding the opposite opinion) in various models and experiments described in \ssref{sec_models}{sec_data} \citep{Abrams:2003,Eriksson:2009}. {\bf (c)} Dynamical rules of a continuous opinion model with pair interactions between an individual $i$ and a neighbor $j$ \citep{chacoma2015opinion}. At time $t$, the opinion $x_{i}(t) \in [0,1]$ changes as $x_{i} \rightarrow x_{i} + I_{ij} (x_{j}-x_{i})$ after an interaction with $j$. The influence factor $I_{ij}$ covers an array of behaviors: keep ($I_{ij}=0$), compromise [$I_{ij} \in (0,1)$] or adopt ($I_{ij}=1$) opinions.}
\label{fig:fig1}
\end{figure}

The simplest assumption of opinion dynamics models is to consider a binary state variable $\lbrace x_{i}(t) \rbrace_{i=1, \dots, N}=0,1$ at time $t$ for each of the $N$ agents in a social network, whose interpretation depends on the context we wish to describe: liberal or conservative, left vs. right, for or against some regulation, etc. (\fref{fig:fig1}). The binary-state assumption simplifies the mathematical treatment of the model, allowing us to perform a thorough study of all possible outcomes of its temporal evolution. When we only have two states, the dynamical rules of the model are encapsulated by the so-called {\it transition rates}, the probabilities per unit time that an individual changes its opinion from one state to the other \citep{Gleeson:2011}. In analogy with epidemic spreading \citep{Satorras:2015}, agents become `infected' ($x = 0 \rightarrow 1$) with rate $F$ and `recover' ($x = 1 \rightarrow 0$) with rate $R$ (\fref{fig:fig1}a). The functional form of the transition rates $F$ and $R$ determines the type of model we have, as well as its stationary states of collective opinion (for examples of transition rates in standard models see \citealp{Gleeson:2013}).

Writing models in terms of the functional form of the transition rates allows us to flexibly explore hypotheses of social interaction in both synthetic and empirical opinion dynamics data (\fref{fig:fig1}b and \sref{subsec_rates_exp}). The simplest hypothesis is to assume a linear dependence of the opinion change rates on the relative fraction of neighbors in the opposite state, i.e., the {\it voter model}, either in the absence \citep{Clifford:1973,Holley:1975} or presence \citep{Kirman:1993,Granovsky:1995} of noise. The linear hypothesis corresponds to a mechanism of blind imitation, where agents copy the opinion of randomly selected neighbors. Non-linear mechanisms have been proposed as generalisations of the voter model for specific applications, like in language \citep{Abrams:2003,Vazquez:2010} and species (spatial) competition \citep{Schweitzer:2009,Vazquez:2008c}, or in more involved copying processes of opinion dynamics \citep{Castellano:2009b,Nyczka:2012,Nyczka:2013,Jedrzejewski:2017,Peralta:2018,Vieira:2018,Vieira:2020}. In the $q-$voter model \citep{Castellano:2009b,Nyczka:2012}, for example, unanimity of opinion in an influence group (of size $q$) is necessary for an individual to imitate its neighbors' opinions. This implies a non-linear dependence of the infection rate $F$ on the fraction $y$ of neighbors in the opposite state, $F(y) \sim y^{q}$. Other functional forms have been proposed, such as the majority-vote model \citep{deOliveira:1992}, where a step function transition rate implies that individuals only copy the majority opinion in their neighborhood. Some variations \citep{Vieira:2018,Vieira:2020} relax the step-wise assumption of the majority-vote model and propose a sigmoid infection rate $F$, intended to model (non-) conformist behavior in the copying of opinions by means of a tunable parameter (\fref{fig:fig1}b).

The transition rates $F$ and $R$ are called up-down symmetric when the dynamical equations of the model are invariant to the state exchange $x = 0 \leftrightarrow 1$. This property further simplifies the mathematical formulation and analysis of many opinion dynamics models \citep{Gleeson:2013,porter2016dynamical}. In a sociopolitical context, for example, this symmetry can be interpreted as a scenario where individuals with opposite opinions behave in the same way (i.e., their response to the influence of people with whom they disagree is the same). This stands in contrast to standard binary models of epidemic spreading or simple contagion [such as the SI and SIS models \citep{Satorras:2015}], which lack up-down symmetry since the biological processes of infection and recovery from a disease are typically different. Another prototypical phenomenon lacking up-down symmetry is complex contagion \citep{granovetter1978threshold,Morris:2000,Montanari:2010}, considering situations where agents have two alternatives and the costs or benefits of the decision depend on how many others have chosen each option, such as in the diffusion of news and rumors, or the adoption of technological innovations \citep{easley2010networks,Karsai:2014,lehmann2018complex,zhang2019empirically}. Similarly to the majority-vote model, threshold-based models of complex contagion \citep{Watts:2002,Guardiola:2002,Llas:2003,dodds2004universal,Centola:2007,Zhongyuan:2015,Karsai:2016} have a step-wise infection rate $F$, meaning agents need a given number or fraction of neighbors in the opposite state to change their own. Other infection rates have been considered in social diffusion problems, in between linear and a step function depending on a few parameters, for example the Hill function \citep{Tuzon:2018}. The recovery rate $R$ can have a different functional shape, though, arguably since the mechanism behind forgetting (or losing interest) in a piece of information, behavior, or product might be different from the one driving their adoption in the first place.

\begin{figure}[t]
\centering
\includegraphics[width=0.46\textwidth]{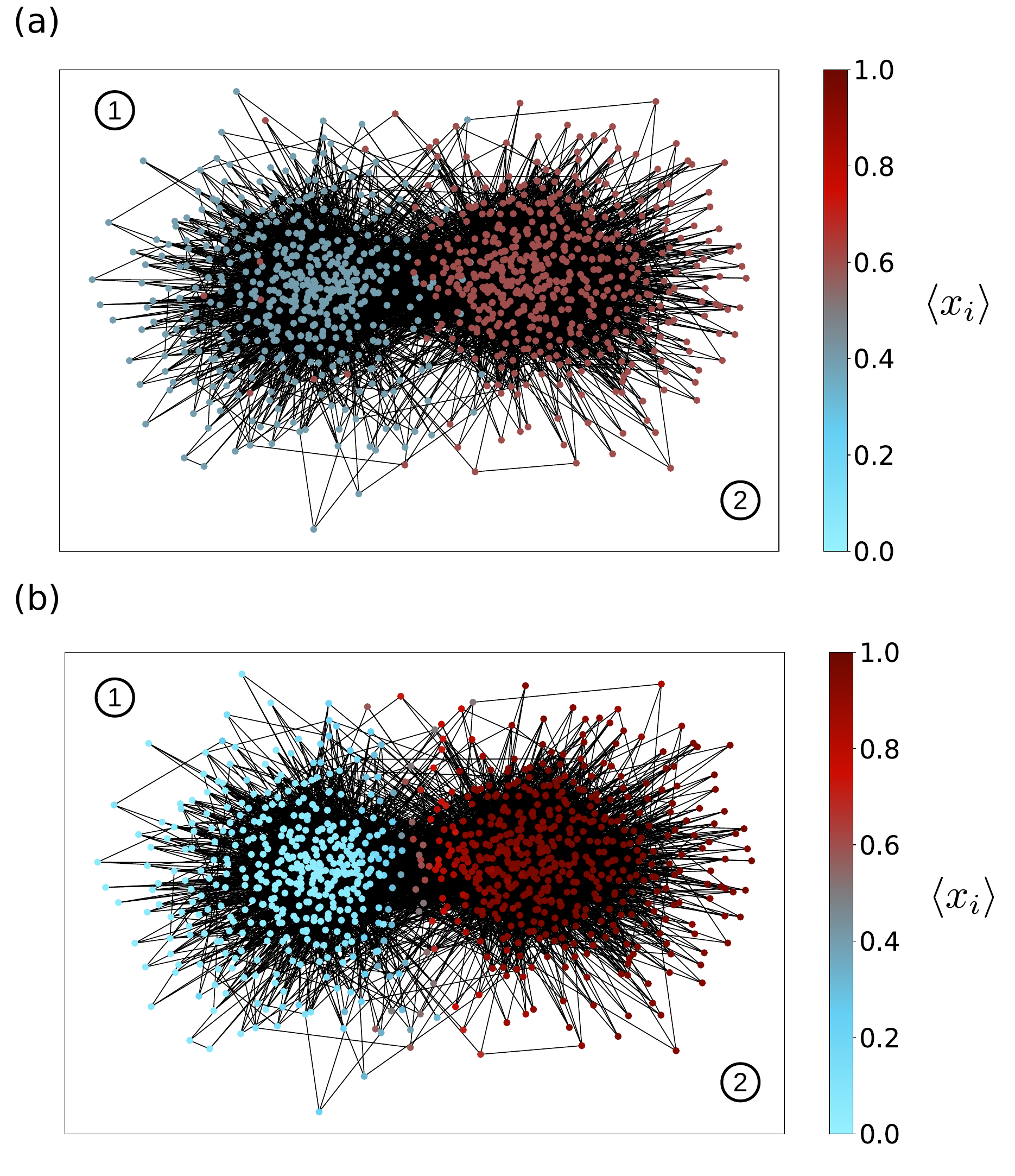}
\includegraphics[width=0.34\textwidth]{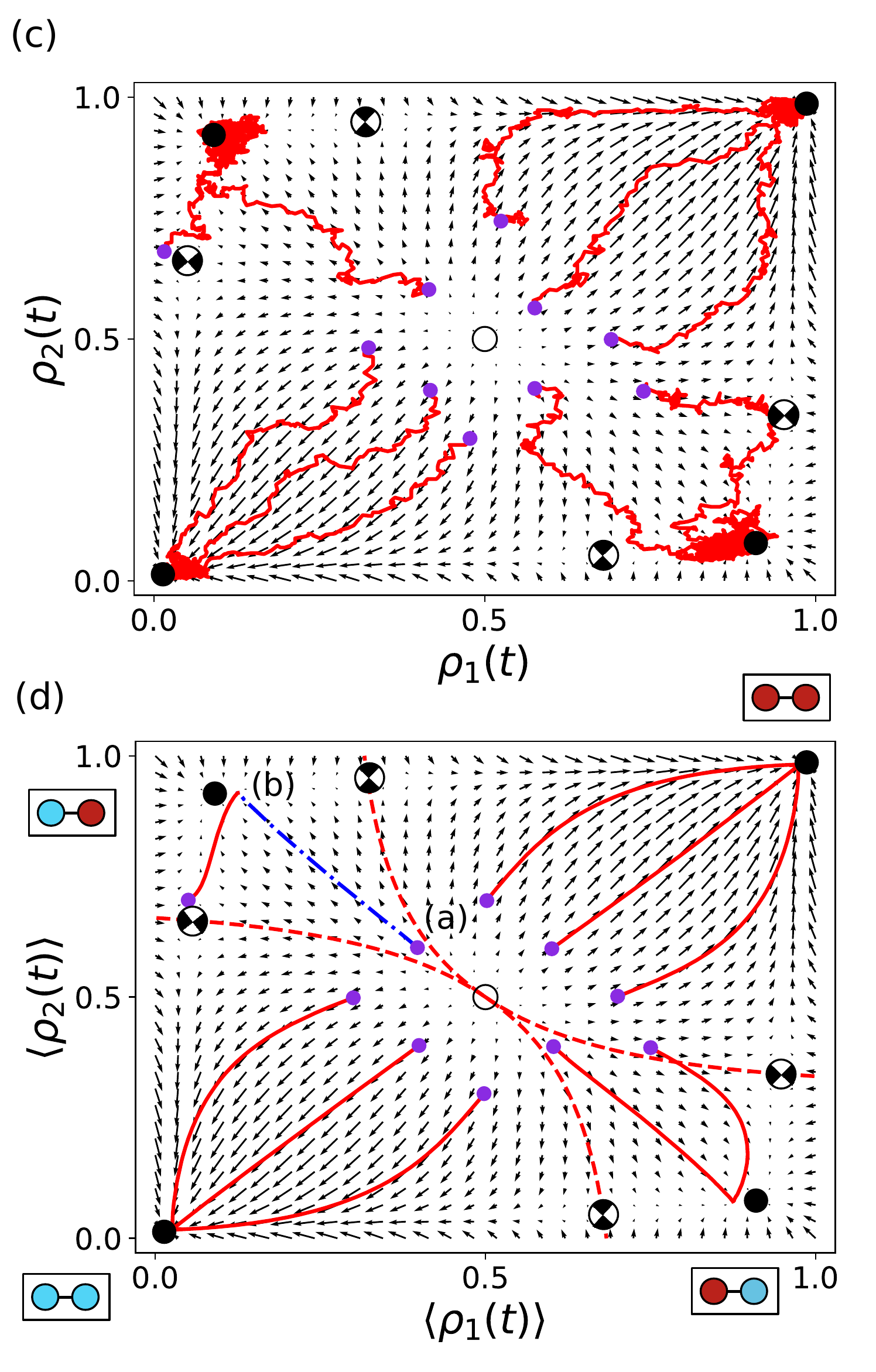}
\caption{\footnotesize {\bf Simulations and approximations of discrete opinion dynamics.} {\bf (a-b)} Numerical simulations of a modified noisy voter model with a conformist type of infection rate (\fref{fig:fig1}b) \citep{Peralta:2021a} on top of an empirical network linking liberal/conservative blogs during the 2004 US elections \citep{Adamic:2005}. Colors denote the average opinion $\langle x_{i} \rangle$ of each individual in the initial (a) and final (b) states of the dynamics. As time goes by, the system becomes more segregated, until two emerging and opposite opinion groups coincide with the community structure underneath. {\bf (c-d)} Vector field of the corresponding mean-field approximation, with $\rho_{1,2}(t)$ the fraction of individuals in group $1,2$ holding opinion $x=1$. Continuous lines denote dynamical trajectories in single numerical simulations (c) and in averages over realizations (d). The dashed line in (d) shows the trajectory from the initial condition in (a) towards the final state in (b). Small circles denote initial conditions; the rest are stable/unstable fixed points. Diagrams in the corners depict the majority opinions inside groups 1 and 2 in the corresponding areas of the vector field (full details in \citealp{Peralta:2021a}).}
\label{fig:fig2}
\end{figure}

There is a large body of literature dedicated to the mathematical description of opinion (and spreading) dynamics on networks at increasing levels of accuracy, from heterogeneous mean field \citep{Sood:2005,Vespignani:2011} and pair approximations \citep{Vazquez:2008a,Pugliese:2009,Vazquez:2010,Mata:2014,Peralta_pair:2018} to approximate master equations \citep{Gleeson:2011,Gleeson:2013,Zhongyuan:2015,Karsai:2016,Unicomb:2018,Peralta:2020,unicomb2021dynamics}. By reducing agent-based dynamics to coupled systems of non-linear differential equations, mean-field and higher-order approximations are useful in intuitively connecting initial conditions, network structure, and model parameters with global states of consensus, coexistence and polarization of opinions. Depending on the model, i.e. the functional form of the transition rates, the temporal evolution and stationary final states of the dynamics predicted at various levels of approximation may coincide or differ (see \citealp{Gleeson:2013} for a thorough analysis). In a variation of the noisy voter model \citep{Peralta:2021a}, for example (a binary opinion model with up-down symmetric rates and a conformist type of infection rate; see Fig. \ref{fig:fig1}b), even an approximation at the mean-field level is capable of recovering numerical simulations of the dynamics over empirical networks with community structure (\fref{fig:fig2}).
    
The generalization of opinion variables to hold more than two states is one attempt at making voter-like models more realistic \citep{Herrerias:2019,Vazquez:2019}. Curiously, binary-state and multi-state versions of the voter model behave in similar ways, with equivalent statistical properties after appropriately re-scaling parameters. This means that a simpler, binary-state approach may be enough, in most situations, to capture opinion dynamics driven by copying mechanisms. Multi-state opinion models are also related to idealized descriptions of cultural dissemination. In the Axelrod model \citep{Axelrod:1997,Klemm:2003}, for instance, each individual is characterized by a finite vector of cultural features, with each feature taking one of $q$ possible values, called cultural traits. Similarly to the voter model, agents copy some of the cultural features of neighbors after interacting with them, but with probability proportional to the fraction of shared common features. Imitation is not `blind' then, as in the voter model, but more likely for already similar individuals, implementing the notion of homophily \citep{mcpherson2001birds,asikainen2020cumulative}. The Axelrod model shows a phase transition from mono- to multi-cultural global states as the number $q$ of cultural traits increases.

We now turn to opinion dynamics models with continuous state variables, keeping in mind 
that a reduction to discrete opinions is often possible \citep{Vazquez:2020,Saintier:2020}. If this reduction captures the phenomenology of the dynamics with enough precision, we can use it to simplify the mathematical treatment of the continuous model with the techniques described above.
    
\subsection{Models with continuous opinions}
\label{sec_continuous_model}
    
A widely studied class of models considers the opinion variables $\lbrace x_{i}(t) \rbrace_{i=1, \dots, N}$ to be real numbers (or vectors, with multiple real components) \citep{stauffer2005sociophysics}. The idea is to represent the opinions of individuals on one or several topics not as a predefined set of possibilities, but as perspectives in a spectrum, with a notion of distance between them. Continuous opinion models might use an (in-)finite interval for their state variables (like $x_{i}(t) \in [0,1]$ in \fref{fig:fig1}c), or even periodic boundary conditions such that extremes of the interval actually represent the same opinion.

In bounded confidence dynamics, also known as the Deffuant model \citep{Deffuant:2000,Weisbuch:2002,lorenz2007continuous}, opinions lie in a finite real interval, $x_{i}(t) \in [0,1]$, and individuals interact pairwise only if their opinions are similar enough ($\vert x_{i}(t) - x_{j}(t) \vert < \varepsilon$). After the interaction, state variables get updated as
\begin{eqnarray}
\label{Deff_eq1}
    x_{i}(t+1)&=&x_{i}(t)+\mu \left[ x_{j}(t) - x_{i}(t) \right], \\
     \label{Deff_eq2}
    x_{j}(t+1)&=&x_{j}(t)+\mu \left[ x_{i}(t) - x_{j}(t) \right],
\end{eqnarray}
that is, opinions become more similar with an amplitude proportional to the original opinion distance, regulated by a parameter $\mu$. Opinion dynamics under bounded confidence have been studied extensively \citep{Castellano:2009,xia2011opinion,Sirbu:2017} [see, e.g., a recent review of its behavior on networks \citep{Fennell:2021} and a generalization to multidimensional opinions \citep{Pedraza:2021}]. Other extensions consider noise as random opinion switching \citep{Pineda:2009, Carro:2013}, algorithmic bias due to content filtering technologies \citep{Sirbu:2019}, and interaction with a medium when modeling conflict in online collaboration \citep{torok2013opinions}. 

\begin{figure}[t]
\centering
\includegraphics[width=0.45\textwidth]{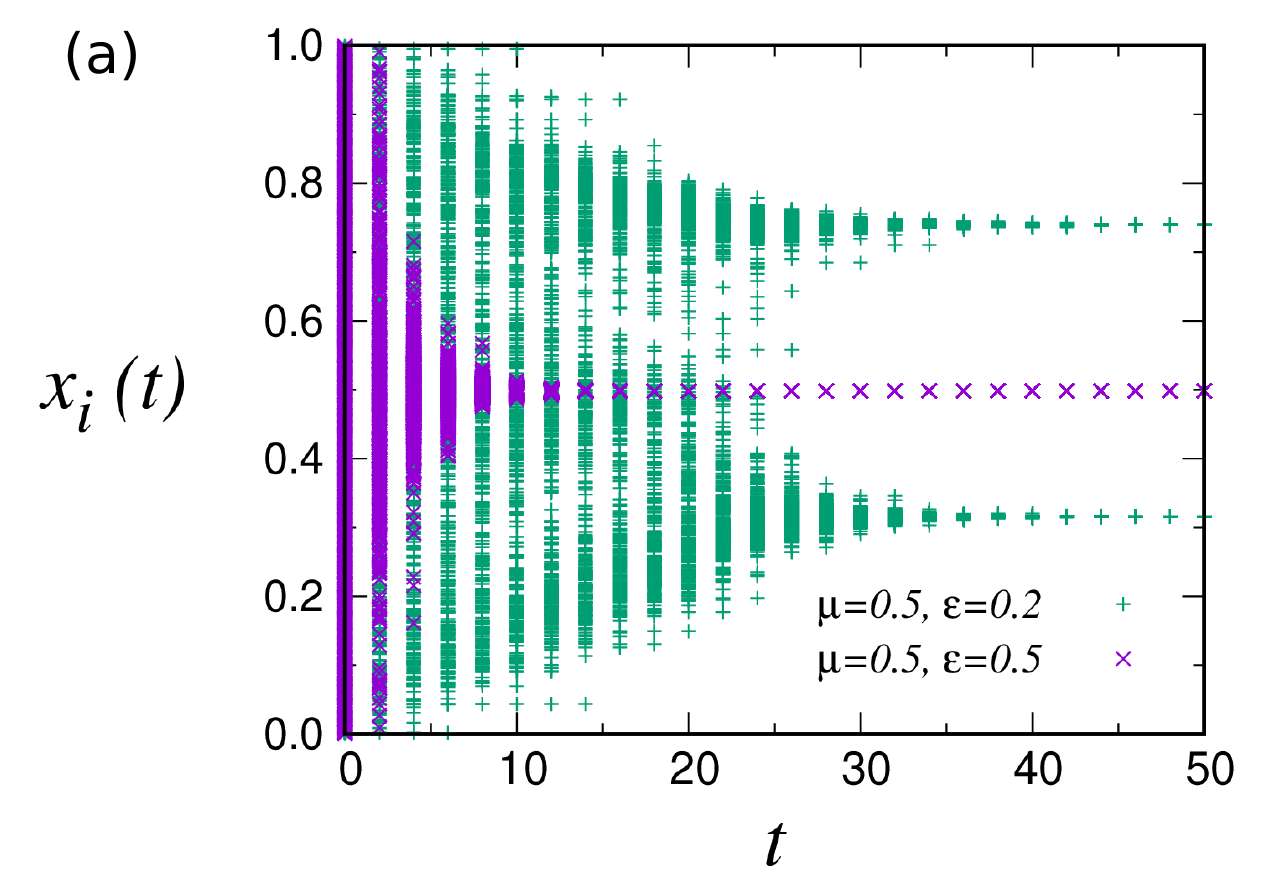}
\includegraphics[width=0.45\textwidth]{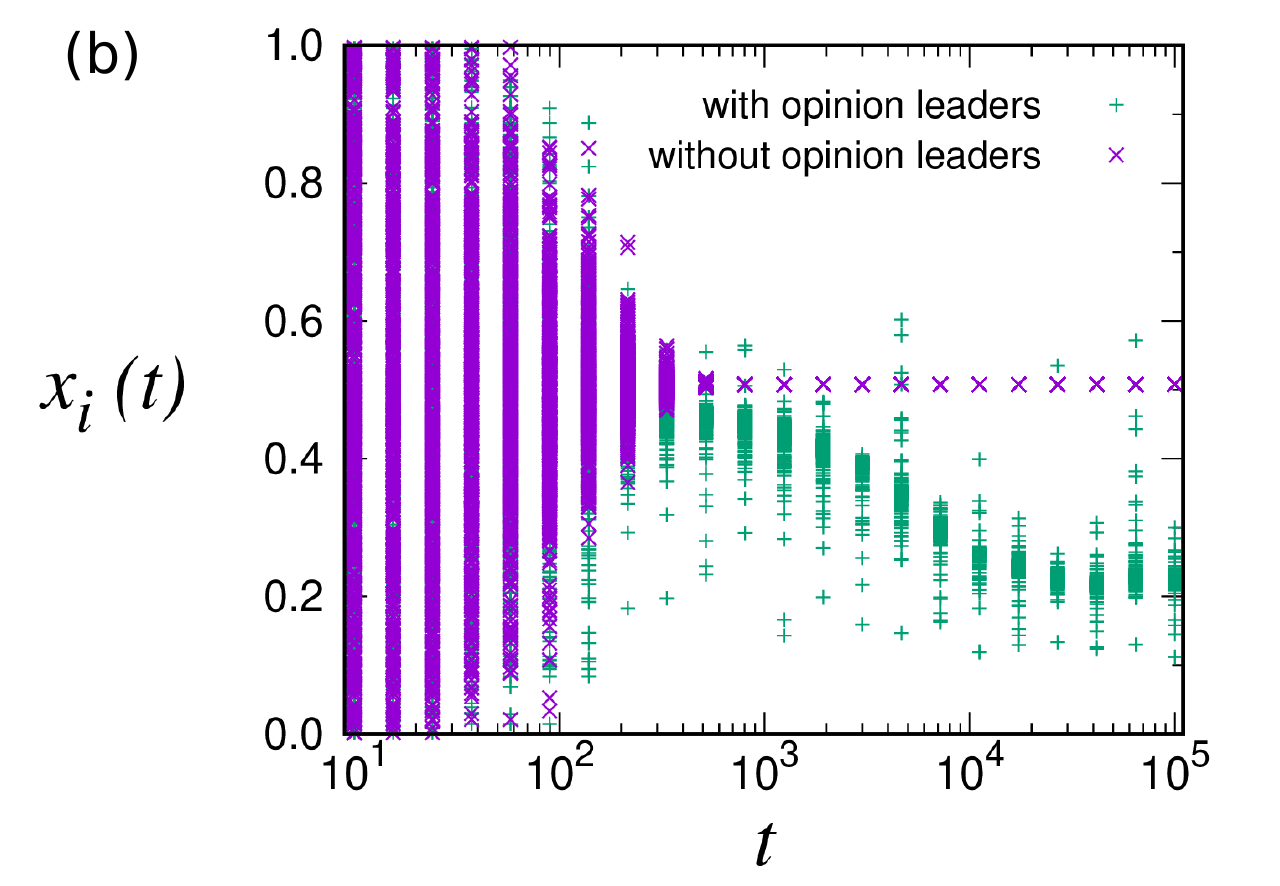}
\caption{\footnotesize {\bf Simulations of continuous opinion dynamics.} Numerical simulations of two continuous opinion models with $N=10^3$ individuals, {\bf (a)} the Deffuant model of \esref{Deff_eq1}{Deff_eq2} \citep{Deffuant:2000}, and {\bf (b)} the bounded confidence model of Chacoma and Zanette \citeyearpar{chacoma2015opinion} (see \fref{fig:fig1}c). We use uniformly random initial conditions in the interval $x_{i} \in [0,1]$ for both cases, and show the time evolution $x_{i}(t)$ of each agent, where the time unit $\Delta t=1$ corresponds to $N$ pairwise interactions. In (a), the strength parameter is $\mu=0.5$ and the bounded confidence parameter is either $\varepsilon=0.2$ (`+' symbols) or $\varepsilon=0.5$ (`x' symbols). In (b) we show two contrasting scenarios: (i) an endogenous bounded confidence dynamics leading to consensus (`x' symbols), and (ii) the case when two opinion leaders are present (`+' symbols). Leaders have fixed opinions $x_{N+1}=0.7$ and $x_{N+2}=0.1$, and interact $\alpha_{1}N=\alpha N$ and $\alpha_{2} N= 4 \alpha N$ times more often with the rest of the agents ($\alpha=10^{-3}$).}
\label{fig:fig3}
\end{figure}

For high enough $\varepsilon$, \esref{Deff_eq1}{Deff_eq2} drive the system towards {\it consensus} (\fref{fig:fig3}a), a global state where all individuals have the same opinion value $x_{i}(\infty)=\tilde{x}$, located around the initial average opinion $\tilde{x} \approx \langle x(0) \rangle$. Below a certain value of $\varepsilon$ we obtain {\it polarization} instead, where individuals belong to one of two opinion groups in the long term (Fig. \ref{fig:fig3}a). For low enough values of $\varepsilon$, we finally obtain a state of {\it fragmentation} with 3 or more stationary opinions. Bounded confidence dynamics are strongly dependent on initial conditions; noise can decrease this dependence and add dispersion to the final distribution of opinions \citep{Carro:2013}. As we describe in \sref{sec_continuous_exp}, bounded confidence is one of the few mechanisms in opinion dynamics with some degree of empirical validation via computerized experiments \citep{chacoma2015opinion}. In a variation of the Deffuant model (see \fref{fig:fig1}c), consensus is avoided in the presence of opinion leaders, agents with fixed opinions that influence the rest of the population (\fref{fig:fig3}b). Both model and experiments show two stages in their dynamics: In the first, opinions evolve rapidly as if leaders were not present, reaching consensus around the average initial opinion. In the second stage, however, there is a slow evolution towards a stationary state dominated by the opinions of leaders. Disagreement between leaders produces a final state with no consensus and thus opinion dispersion. Similar studies have also considered individuals with fixed opinions, referred to as zealots, in discrete opinion models \citep{Kononovicius:2014,Khalil:2018}.

Other models with continuous variables have been proposed, both in the field of opinion dynamics and in other contexts such as animal collective behavior \citep{Vicsek:1995}. In flocking models, the variable $x_{i}(t)$ does not represent an opinion but the angular direction of movement. The analogy between flocking and opinion models goes beyond conceptual similarity, though; the noisy voter model turns out to be a good theoretical descriptor for the collective behavior of both ants and humans \citep{Kirman:1993} (see \sref{sec_data}).

In the DeGroot model \citep{DeGroot:1974,Dandekar:2013}, a paradigmatic dynamics of continuous opinions, states are a weighted average over opinions in the network neighborhood. The update rule is usually asynchronous, meaning changes depend only on opinion states in the previous step. The opinion $x_{i}(t) \in [0,1]$ evolves according to
\begin{equation}
\label{DeGroot_eq}
        x_{i}(t+1) = \sum_{j=1}^{N}  p_{ij} x_{j}(t),  
\end{equation}
where $p_{ij}$ are normalized weights with $\sum_{j=1}^{N} p_{ij} = 1$. The DeGroot model reduces to an algebraic problem \citep{DeGroot:1974}, with dynamics dependent on the topology of interactions~\citep{Jackson:2008}. Iterating Eq. (\ref{DeGroot_eq}), the system ends up in a consensus state without opinion dispersion (if certain conditions on the matrix ${p_{ij}}$ are fulfilled), as in the Deffuant model with high $\varepsilon$. Highly studied variations of the DeGroot model \citep{Dandekar:2013} are, for example, the Friedkin-Johnsen model \citep{Friedkin:1990}, in which a term dependent on the initial condition is added in the left hand side of Eq. (\ref{DeGroot_eq}); and the Hegselmann-Krause model \citep{Hegselmann:2002}, in which the sum in Eq. (\ref{DeGroot_eq}) is restricted with a bounded confidence condition. An analogy in the context of flocking is the Vicsek model \citep{Vicsek:1995}, where agents move in space with certain velocity, and their direction of movement aligns to the average direction over a restricted neighborhood (based on spatial proximity). There are similarities in the influence mechanisms proposed by the DeGroot and Vicsek models, in the sense that both consider the average state of the neighbors. There are crucial differences as well; in the flocking model agents move in space and the neighborhood constantly changes in time.

In other flocking models \citep{Vazquez:2019,Jhawar:2020}, similar to multi-state voter models \citep{Herrerias:2019}, the opinion-copying mechanism is based on pairwise influence:
\begin{equation}
\label{pair_inf}
    x_{i} + x_{j \neq i} \rightarrow 2 x_{j},
\end{equation}
that is, individual $i$ copies the state of neighbor $j$ with a certain rate. Higher-order interactions can also be considered, such as ternary,  
\begin{equation}
\label{group_inf}
    x_{i} + x_{j \neq i} + x_{k \neq j \neq i} \rightarrow \begin{cases}
    2 x_{j} + x_{k},\\
     x_{j} + 2 x_{k},
\end{cases}  
\end{equation}
where agent $i$ copies the state of a neighbor, either $j$ or $k$, only if the distance between them ($\vert x_{j}-x_{k} \vert$) is smaller than between $i$ and $j$ or $i$ and $k$ (i.e. the continuous version of the $q-$voter model with $q=2$). Flocking and multi-state voter models have been compared to experiments on animal movement \citep{Buhl:2006,Jhawar:2020}. Multi-state voter dynamics differ from flocking (Vicsek model) in that the state of single neighbors is copied, the adopting mechanism follows \fref{fig:fig1}c, and there is no gradual approaching of opinions like in bounded confidence models, or opinion averaging like in the DeGroot or Vicsek models. The voter model further differs in some of its statistical properties and dependence on system size \citep{Vazquez:2021}.

Several models, originally built to study other types of collective behavior, have been interpreted as dynamics of opinion, usually in the form of coupled non-linear differential equations for appropriate state variables. An opinion model for polarization in multidimensional topic spaces has recently been proposed by \citet{Baumann:2021}, in which the associated equation (in the case of a single topic) reads
\begin{equation}
\label{multi_eq}
    \frac{d x_{i}}{d t} = - x_{i} + \sum_{j=1}^{N} A_{ij} \tanh\left[ \alpha x_{j} \right],
\end{equation}
where opinions are unbounded variables $x_{i}(t) \in (-\infty, +\infty)$, $A_{ij}$ is the connectivity matrix, and $\alpha$ is a measure of the `controversy' of the
topic. Opinions tend to converge to the center $x_{i}=0$ [due to the first term in \eref{multi_eq}], unless the influence of the neighborhood pushes it to a different fixed point [determined by the second term in \eref{multi_eq}]. The parameter $\alpha$ tunes the sensibility to the opinions of acquaintances, and the functional form of influence is an hyperbolic tangent, which saturates to $\pm 1$ in the limit $x_{j} \rightarrow \pm \infty$. Collective states representing uncorrelated polarization, ideology and consensus are compared to empirical data coming from surveys in the US \citep{Baumann:2021}. \eref{multi_eq} is similar to the widely studied Kuramoto synchronization model \citep{Kuramoto:1975},
\begin{equation}
\label{Kuramoto_eq}
    \frac{d x_{i}}{d t} = \omega_{i} + \sum_{j=1}^{N} A_{ij} \sin\left[ \pi \left(x_{i}-x_{j}\right) \right].
\end{equation}
In this case the differential equations represent a set of coupled oscillators of frequency $\omega_{i}$, with the state variables $x_{i}(t) \in [0,1]$ an angle or phase with periodic boundary conditions. The Kuramoto model has been applied to applause dynamics \citep{Neda:2000}, where state variables are `phases' and the global variable corresponds to the sound intensity of applause. The coupling is a sinusoidal function of the phase difference $x_{i}-x_{j}$, which saturates to $\pm 1$ as in \eref{multi_eq}, this time with bounded variables.

\section{Data in opinion dynamics}
\label{sec_data}

Models of opinion dynamics can be validated with empirical data in several ways. One is to compare the aggregate behaviour of voting and similar opinion models \citep{merrill1999unified,easley2010networks} with data on population-wide polls and elections \citep{bernardes2002election,Fortunato:2007,braha2017voting}, or with user-generated content tailored by online algorithms \citep{liu2012sentiment,bakshy2015exposure}. A comparison with observational data is often not enough to detect causal relations between quantities of interest, but can indicate generic patterns of behavior across seemingly different social systems, and pinpoint what mechanisms of interaction might lead to a particular distribution of opinions in the real world. The development of computerized social network experiments can further probe how exactly individuals influence each other in sociopolitical contexts \citep{bail2018exposure,stewart2019information,balietti2021reducing}, for example by providing data on transition rates between opinion states.

\subsection{Election and poll data}
\label{subsec_elections}

There is an array of studies in the literature exploring statistical regularities in election data, particularly how votes (understood as favorable opinions towards political candidates) are distributed in the population. In the 1998 Brazilian elections, for example, the number of candidates $N(v)$ with $v$ votes exhibited approximate power-law behavior, $N(v) \approx v^{-1}$ \citep{Filho:1999}. Several models have tried to explain this scaling property as the result of an opinion spreading process \citep{bernardes2002election,Chatterjee:2013}. Later on, the same trend was observed in additional datasets \citep{Fortunato:2007,Chatterjee:2013}, with the lognormal distribution as a good fit for several countries and electoral periods. It is necessary, however, to introduce as rescaling factor the average number $v_0$ of votes of all candidates in the same party, in order to collapse the various curves $N(v)$ vs. $v/v_{0}$ into a single universal (lognormal) function. A simple model \citep{Fortunato:2007} is able to emulate these observations, where candidates are the roots of a tree-like network of interactions among voters, and there is a power-law distribution in the number of contacts. Opinions then spread from the roots of the network (the candidates) to its leaves (the voters), following a simple contagion process. 

Similar studies have focused on the distribution of vote shares across polling stations or regions of a country \citep{Araripe:2006,Klimek:2012}. The voter model is actually able to reproduce data on vote-share distributions at county level in the 1980--2012 US presidential elections \citep{Gracia:2014}. The noisy voter model, which produces a beta distribution of vote shares, explains data on the Lithuanian parliamentary elections of 1992, 2008, and 2012 \citep{Kononovicius:2017,Kononovicius:2018}. Some studies on modelling electoral processes have used Wikipedia data in an attempt to predict electoral results \citep{Yasseri:2016}. \citet{Phan:2019} have fitted a binary-opinion model to poll data in the 2012--2016 US presidential elections with the intent of revealing underlying social influence mechanisms.

\subsection{Data on discrete opinions}
\label{subsec_rates_exp}

Controlled sociological experiments are a rich source of empirical data to validate idealized models of opinion dynamics, particularly to identify which influence mechanisms are the most relevant in emulating human social behavior \citep{Mercier:2019}. In these experiments, individuals choose between various options after being exposed (or not) to information coming from other subjects or the researchers themselves. A typical quantity to be measured is the percentage of subjects that change their perspective (or opinion) as a function of various controlled variables, such as: the absolute and relative number of people holding that opinion; their competence or opinion strength (i.e. status, power, or ability); the prior beliefs of the individual, etc.

As described in \sref{sec_discrete_models}, a key ingredient of binary opinion models is the infection rate $F$. Thus, a reasonable output of experiments on social influence is information on its functional form. There is some debate among social psychologists on the shape of $F$ as a function of the relative number of individuals holding an opinion, $y=m/k$ (see \fref{fig:fig1}b). Social impact theory \citep{Latane:1981b,holyst2001social,Eriksson:2009} suggests an inverted `S' shape, such that the minority opinion has relatively higher influence [a sub-linear trend, $F(y) \sim y^{\alpha}$ with $\alpha<1$, like in the language model \citep{Abrams:2003} and the nonconformist case in \fref{fig:fig1}b]. This is known as the principle of marginally decreasing impact in social impact theory: if an individual is exposed to the opinions of its acquaintances sequentially, each additional opinion is less influential than previous ones. An `S' shape in $F$, implying that the minority opinion has relatively lower influence [a super-linear trend, $F(y) \sim y^{\alpha}$ with $\alpha>1$], has also been proposed \citep{Tanford:1984,Morgan:2012} as the so-called `conformist' transmission of opinions \citep{Boyd:1996,Henrich:1998}. In this case, an individual needs opinion unanimity (agreement) between its information sources in order to change its own. In Ref. \citep{Eguiluz:2015} a Bayesian theory applied to experimental data shows a conformist type of social behavior, although it is difficult to distinguish from a linear trend, i.e., $F(y) \sim y^{\alpha}$ with $\alpha \approx 1$.



In non-linear voter models \citep{Vazquez:2010,Peralta:2018,Vieira:2020,Peralta:2021a}, an inverted or straight `S' shape in the infection rate $F$ leads either to coexistence or consensus and polarization of opinions, respectively. Given the variability in experimental setups and their sometimes conflicting results \citep{Mercier:2019}, we can expect empirical data on $F$ to depend on the social context studied and the particularities of each experiment. In social impact experiments, for example, individuals do not have a strong opinion on the topic before being influenced by social information \citep{Eriksson:2009}, while in conformist bias experiments subjects do have an existing opinion based on prior and/or sensory information \citep{Morgan:2012}. The latter scenario might be more comparable to models of opinion dynamics, as consensus or opinion polarization are observed more often in real-world phenomena like language competition \citep{Abrams:2003} and electoral processes \citep{Peralta:2021a}.

Models of discrete opinions have been compared to empirical data across various disciplines. A version of the non-linear voter model (known as the Abrams-Strogatz model) has been used to explain the evolution of the number of speakers in competing languages \citep{Abrams:2003,Mira:2005} and businesses \citep{Legara:2009}. As mentioned previously, variations of the noisy voter model have been applied to electoral data \citep{Gracia:2014,Kononovicius:2017}. In one of its most interdisciplinary applications, known as the Kirman model \citep{Kirman:1993}, the noisy voter model has been proposed as the mechanism behind price formation in financial markets \citep{Alfarano:2005,Alfarano:2008,Alfarano:2009}, people choosing among restaurants \citep{Beckers:1990,Becker:1991}, and ants foraging for food \citep{Deneubourg:1987,Pasteels:1987}.

\subsection{Data on continuous opinions}
\label{sec_continuous_exp}

Similar experiments to the ones described above have been run within a continuous opinion setup \citep{Moussaid:2013,chacoma2015opinion}. In the spirit of bounded confidence models, subjects state their opinion and confidence in it, engaging in this process over several rounds. The experiment of \citet{chacoma2015opinion} identifies three behavioral categories based on how individuals influence each other: (i) {\it keep}, where the opinion of a person remains unchanged; (ii) {\it adopt}, where the subject copies a reference opinion; and (iii) {\it compromise}, where the individual approaches the reference opinion (see \fref{fig:fig1}c). Results indicate that the most common behavior is {\it keep}, followed by {\it compromise} and finally {\it adopt}. The chosen behavior depends on the difference in confidence between the individual and reference opinions. The {\it keep} behavior is more probable when the subject is more confident than the reference; the {\it compromise} and {\it adopt} behaviors are more common in the opposite case. Curiously, there is no significant correlation between the opinion difference and the class of influence behavior. Thus, a metric comparison with reference opinions has a negligible effect on how agents respond to influence, in contrast to some of the continuous opinion models discussed in \sref{sec_continuous_model}.

A recent opinion dynamics experiment \citep{vande2016modelling} has also measured how a set of $N=861$ individuals influence each other [i.e. the factor $I_{ij}$ in \fref{fig:fig1}c, or $\mu$ in \esref{Deff_eq1}{Deff_eq2}]. The distribution of types of influence in the population is comparable to the one reported by \citet{chacoma2015opinion}. A simple model of consensus formation, based on pairwise interactions, is then used to predict the evolution of opinions observed in the empirical data. Beyond these experiments, the dynamics of other continuous opinion models based on \eref{multi_eq} have been compared against polling data on various topics \citep{Baumann:2021}. Additionaly, a version of the Friedkin-Johnsen model has been applied to study consensus formation processes in the 2015 Paris Agreement on climate change \citep{Bernardo:2021}. In the context of animal collective phenomena, a continuous version of the voter model [\eref{pair_inf}] seems to be a more suitable model for flocking than the Vicsek model \citep{Jhawar:2020}. By recording the collective movement of fishes, and then comparing it to the dynamics of the voter model (with both pairwise and higher-order interactions), the authors conclude that the voter model with pairwise interactions is in close agreement with experimental results.

\section{Towards more realistic models of opinion dynamics}
\label{sec_structural}

Idealized models of opinions dynamics are useful as proofs of concept to test the consistency of descriptive theories of collective behavior, and to explore the potential outcomes of sociopolitical scenarios based on simplified hypotheses of human interaction \citep{holme2015mechanistic}. Their ability to forecast real-world social systems is, however, limited [and for some phenomena, perhaps nonexistent \citep{sayer1999realism}]. In an attempt to increase the predictive power of the simple models described above, plenty of effort has gone into considering more realistic aspects of the structure and dynamics of opinion sharing.

The models in \sref{sec_models} consider individuals as nodes in a potentially heterogeneous but static social network of acquaintances, where links are persistent in time (or change in a longer time scale relative to the dynamics of opinion). Since connections do not change, highly connected groups of individuals will tend to have similar opinions \citep{granovetter1973strength} [see \fref{fig:fig2} for an example, where noisy voter dynamics in clustered networks lead to a polarized state where community structure and opinion segregation coincide \citep{Peralta:2021a}]. Previous works have considered the role of modular structure on opinion and spreading dynamics, such as the voter model on a two-clique network with a fraction of randomly distributed inter-clique links \citep{Sood:2008,Masuda:2014,Gastner:2019}, and the Watts threshold model of social contagion \citep{Watts:2002} over modular random networks \citep{Gleeson:2008}. The other extreme is a situation where social ties may change but opinions remain fixed [becoming attributes that generate identity within a group \citep{mcpherson2001birds}]. Links are more likely to be rewired between people with similar opinions, due to homophilic or assortative mixing mechanisms \citep{Sooderberg:2002,Newman:2003,Boguna:2004b,toivonen2009comparative,Murase:2019}. The dynamic interplay between homophily and the formation of triangles leads to highly clustered networks, or networks with a core-periphery structure \citep{asikainen2020cumulative}.

Real-world opinion dynamics are likely somewhere in between, with both opinions and social network changing at the same time. Coevolving or adaptive opinion models consider this scenario, typically via a parameter (called plasticity or rewiring rate) measuring the probability of an opinion or link update during a time unit \citep{gross2008adaptive,zschaler2012adaptive}. Coevolving dynamics have been extensively studied for both discrete and continuous opinions \citep{Holme:2006,Nardini:2008,kozma2008consensus,Kozma:2008,iniguez2009opinion,Toruniewska:2017}, in the Axelrod model of cultural dissemination \citep{Vazquez:2007} and for language competition models \citep{Carro:2016b}, among many more. Other studies consider additional features of opinion dynamics like noise \citep{Diakonova:2015}, non-linear copying mechanisms \citep{Raducha:2018,Raducha:2020,Jedrzejewski:2020}, and the possibility that individuals conceal their opinions from others \citep{iniguez2014effects,barrio2015dynamics}. When social ties between agents with similar opinions are favored, adaptive dynamics self-organize into heterogeneous networks where groups of individuals sharing attributes are structurally distinguishable from each other, leading, e.g., to fragmentation--consensus transitions as a function of the rewiring rate \citep{Holme:2006,iniguez2009opinion}.

A relevant aspect of more realistic opinion models is multilayer network structure \citep{Kivela:2014}. The idea behind is that people do not discuss and influence each other within a single context, but across an array of layers of interaction, each with varying structure and potentially different rules for opinion spreading \citep{Diakonova:2014}. Network layers can be interpreted as topics or perspectives in a multifaceted discussion, or as social contexts where people behave differently (due to cultural norms and other constraints), such as work, family, and friend circles, or various online social platforms. A key issue addressed by multilayer opinion models is the irreducibility of the dynamics \citep{Diakonova:2016,Min:2019}, i.e. in which cases, and for what parameter values, the phenomenology of multilayer dynamics can be reproduced in the aggregated network (where layers are not differentiated). A similar notion drives opinion models with mixed rules \citep{Kurmyshev:2011,Juul:2019}, where sets of individuals follow different mechanisms of interaction. While in multilayer models heterogeneity comes from the layer structure, in mixed rule models it arises from the response of individuals themselves, leading to insight on the way antiestablishment choices might fare in competitions between political candidates, products, etc.

Many other models are out there. Some consider memory effects like `inertia' \citep{Schweitzer:2009} or `aging' \citep{Gracia:2011,Perez:2016,Artime:2017b,Artime:2018a,Peralta:2019,Peralta:2020b} of agents, where the probability of changing opinion decreases with the time holding the same state (an individual's `age'), similar to the inflexibility notion considered in \citep{Martins:2013}. This mechanism generates bursty signals \citep{Gracia:2011}, i.e. broad inter-event time distributions like the ones often observed in empirical human communication data \citep{Candia:2008,Wu:2010,karsai2012universal}. In aging models burstiness emerges due to memory effects, but several studies also include it as an exogenous, fixed factor in the dynamics \citep{Takaguchi:2011,Hoffmann:2012,Min:2013,unicomb2021dynamics}, typically within the framework of temporal networks \citep{holme2012temporal,holme2015modern}.

\section{Outlook}
\label{sec_conclusions}

Opinions are key to how we perceive the world and those around us, shaping our actions and connections with peers. They are an essential part of the political and social discourse, leading to everything from worldwide cooperation to polarization and strife. Together with epidemic and information spreading, modeling opinion dynamics is an archetypal example of how interconnected decisions of individuals lead to emergent collective behaviour at the level of society. In epidemic modelling, microscopic rules and parameter values are informed by how biological pathogens work, while online communication datasets guide the development of information spreading models. Data in opinions is more sparse, typically coming from detailed but small sociological surveys, or aggregated over population and time in polls and elections. This limitation has led to an explosion in opinion dynamics models without a counterpart in empirical validation. Already a decade ago, reviews on the topic noticed such mismatch \citep{Castellano:2009,sobkowicz2009modelling,Sirbu:2017}; to this day, relatively few empirical studies exist. Part of the reason is perhaps the prevalence of the statistical physics community in the field, accustomed to a wide array of experimental results and fundamental laws over which to carry independent theoretical work. But part is the inherent value in the idealized, mechanistic modelling of opinion dynamics, allowing us to probe scenarios, test descriptive theories for consistency, and explore emergent phenomena \citep{holme2015mechanistic,gonzalez2013social}.

A promising hint in closing this gap between opinion models and data is the rise of computerized social experiments, both in academia and the industry or performed in close collaboration with the general population \citep{sagarra2016citizen}, together with the automated analysis of large datasets of user-generated content online \citep{liu2012sentiment}. Experiments of opinion dynamics in controlled settings may allow to detect causal relations between variables of interest, and even select among competing mechanisms of interaction, such as in the recent observation that exposure to opposing political views on social media can either increase \citep{bail2018exposure} or decrease \citep{balietti2021reducing} polarization. This rise in experimental studies will have to go together with the development of tools to automatically infer microscopic rules of behavior from noisy, incomplete data \citep{chen2019flexible}. The statistical inference of model rules from empirical opinion dynamics can benefit from tools like approximate Bayesian computation~\citep{gutmann2016bayesian}, network-aware nonparametric inference~\citep{peixoto2019network}, and node state detection in temporal data~\citep{masuda2019detecting}.

The continuing relevance of opinion models is particularly clear given the ubiquity and fast growth of algorithm-driven social networking platforms \citep{lazer2009life,conte2012manifesto}. Sharing opinions is no longer mediated by social processes only, such as homophily and structural constraints \citep{asikainen2020cumulative}, but by artificial algorithms filtering the content available to users, based on hidden rules aimed at maximizing platform use \citep{bozdag2013bias,nikolov2018quantifying}. Models of opinion dynamics help clarify the interplay between human behavior and artificial intelligence, hinting at how algorithmic bias can promote echo chambers and opinion polarization online \citep{santos2021link}, or instead help disseminate information across structural divides, depending on the mechanisms of interaction assumed by the models \citep{Sirbu:2019,Perra:2019,Peralta:2021a,peralta2021opinion}. Online social networks are expected to provide empirical background for studying opinion dynamics by means of natural language processing~\citep{Willaert:2020}

Modeling opinion dynamics in social networks shows the potential of an interdisciplinary approach to understand the emergence of collective behaviour from the actions and interactions of individuals. How, despite the multifaceted complexity of people's psychology and the many sociopolitical contexts humans engage in, we can still link individual preferences to global states of consensus and polarization of opinions. Together with a more stringent empirical validation and with a systematic statistical identification of fundamental mechanisms, the field of opinion dynamics may advance enough to meaningfully contribute to the conversation of what interventions might steer the social discourse on a specific path. Eventually, it might establish a sound theoretical ground for computational social science efforts in public policy \citep{lazer2020computational}, an ethical development of algorithms behind online social platforms \citep{wagner2021measuring}, and a source of intuition on how to better communicate with each other in our highly interconnected, but ever more fractured world.

\section*{Further reading}

{\footnotesize
\begin{itemize}
\item Castellano, C., Fortunato, S., and Loreto, V. (2009a). Statistical physics of social dynamics. {\it Rev. Mod. Phys.}, 81:591–646.

\item Holme, P. and Liljeros, F. (2015). Mechanistic models in computational social science. {\it Front. Phys.}, 3:78.

\item S{\^i}rbu, A., Loreto, V., Servedio, V. D. P., and Tria, F. (2016). Opinion dynamics: Models, extensions and external effects. In
Loreto et al., V., editor, {\it Participatory Sensing, Opinions and Collective Awareness}, pages 363–401. Springer.

\item Wagner, C., Strohmaier, M., Olteanu, A., Kıcıman, E., Contractor, N., and Eliassi-Rad, T. (2021). Measuring algorithmically
infused societies. {\it Nature}, 595(7866):197–204.

\item Xia, H., Wang, H., and Xuan, Z. (2011). Opinion dynamics: A multidisciplinary review and perspective on future research.
{\it IJKSS}, 2(4):72–91.

\item J\polhk{e}drzejewski, A.  and Sznajd-Weron, K. (2019). Statistical Physics Of Opinion Formation: Is it a SPOOF? {\it C. R. Phys.}, 20(4):244-261.
\end{itemize}
}

\section*{Acknowledgments}

{\footnotesize
We acknowledge support from AFOSR (grant FA8655-20-1-7020) and EU H2020 Humane AI-net (grant \#952026). J.K. is grateful for support by project ERC DYNASNET Synergy (grant \#810115), and G.I. by CIVICA project `European Polarisation Observatory' (EPO). J.K. and G.I. acknowledge support from CHIST-ERA-19-XAI-010 project SAI (grant FWF I 5205-N).
}

\setlength{\bibsep}{4pt}
{\scriptsize \bibliographystyle{apalike}
\bibliography{references}
}

\end{document}